# Transport of hydrogen isotopes through interlayer spacing in van der Waals crystals

S. Hu[1], K. Gopinadhan[2], A. Rakowski[3], M. Neek-Amal[1,4], T. Heine[5], I. V. Grigorieva[2], S. J. Haigh[3], F. M. Peeters[4], A. K. Geim[1,2], M. Lozada-Hidalgo[1,2]

[1]National Graphene Institute, The University of Manchester, Manchester M13 9PL, UK
[2]School of Physics and Astronomy, The University of Manchester, Manchester M13 9PL, UK
[3]School of Materials, The University of Manchester, Manchester M13 9PL, UK
[4]Departement Fysica, Universiteit Antwerpen, Groenenborgerlaan 171, B-2020 Antwerpen, Belgium
[5]Wilhelm-Ostwald-Institut für Physikalische und Theoretische Chemie, Universität Leipzig, Linnéstr. 2, D-04103 Leipzig, Germany

**Atoms start behaving as waves rather than classical particles if confined in spaces commensurate with their de Broglie wavelength. At room temperature this length is only about one angstrom even for the lightest atom, hydrogen. This restricts quantum-confinement phenomena for atomic species to the realm of very low temperatures[1-5]. Here we show that van der Waals gaps between atomic planes of layered crystals provide angstrom-size channels that make quantum confinement of protons apparent even at room temperature. Our transport measurements show that thermal protons experience a notably higher barrier than deuterons when entering van der Waals gaps in hexagonal boron nitride and molybdenum disulfide. This is attributed to the difference in de Broglie wavelength of the isotopes. Once inside the crystals, transport of both isotopes can be described by classical diffusion, albeit with unexpectedly fast rates, comparable to that of protons in water. The demonstrated angstrom-size channels can be exploited for further studies of atomistic quantum confinement and, if the technology can be scaled up, for sieving hydrogen isotopes.**

At thermal energies, the de Broglie wavelength is given by $\lambda_B = h/\sqrt{3mkT}$ where $h$ is the Planck constant, $m$ the particle mass, $k$ the Boltzmann constant and $T$ the temperature[1]. This yields, for example, $\lambda_B^H \approx 1.45$ Å and $\lambda_B^D \approx 1.02$ Å for protons (H) and deuterons (D) at 300 K, respectively. With decreasing $T$, $\lambda_B$ grows and reaches the micrometer range at μK temperatures. Such long $\lambda_B$ have been exploited in matter-wave experiments with ultra-cold atoms[1-3]. At more accessible cryogenic temperatures, quantum-confinement effects[4,5] can also play a role in the adsorption[6-9] and/or diffusion[10,11] of $H_2$ and $D_2$ molecules inside sub-nanometer pores of materials such as zeolites, carbon molecular sieves and metal-organic-frameworks. These isotope effects, collectively known as quantum sieving[4,5], arise from (physical) spatial confinement of atoms and molecules, which distinguishes them from chemical isotope effects that arise due to a difference in vibrational energies of isotopes within molecules[12].

In this Letter, we investigate whether van der Waals (vdW) gaps between atomic planes of layered crystals can be used as naturally occurring angstrom-scale channels for proton transport and whether they exhibit quantum sieving. The interlayer spacing in vdW crystals is accurately determined by X-ray analysis[13]. It is ~3.34 Å for both hexagonal boron nitride (hBN) and graphite, and ~6.15 Å for molybdenum disulfide ($MoS_2$). These distances are notably larger than, for example, the above $\lambda_B$ for hydrogen isotopes. However, the space between the atomic planes is densely filled with electron clouds surrounding the constituent atoms, which do not allow the planes to come any closer. It remains unknown whether there is any space left to allow interlayer permeation of atomic species including protons and deuterons at thermal ($kT$) energies. The interlayer transport has been studied in channeling experiments using accelerated protons and other particles[14,15]. But the energies of these particles, typically in the MeV range, translate into $\lambda_B$ orders of magnitude smaller than the interlayer spacing, so quantum confinement effects play no role in this case. The possibility of thermal-proton transport through vdW gaps has not been investigated so far.

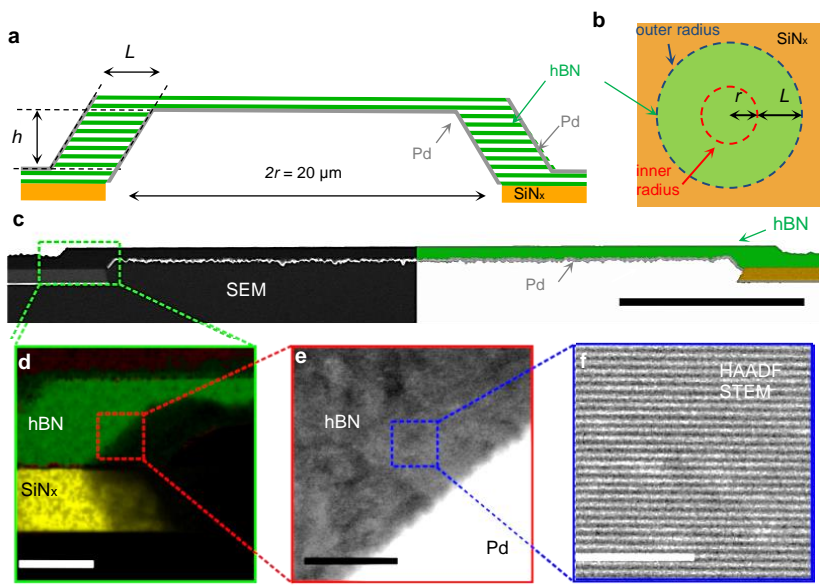

**Figure 1 | Devices for testing interlayer permeation of protons. a,b**, Their schematics. Cross-sectional (a) and top (b) views. A vdW crystal was suspended over 20 μm holes in a $SiN_x$ membrane to separate two compartments so that no gas transport could occur between them. The crystal was then plasma-etched to expose interlayer channels connecting the compartments as shown in **a**. The channels' entries and exits were coated with a thin Pd layer. **c**, Scanning electron microscope image of a cross-section of one of our hBN devices. Scale bar, 5 μm. The right half of the image is false colored to indicate different materials. In the unmodified (left) half of the image, the space above the assembly appears in white because a Pt layer was deposited on top, which was required to make the cross-sectional slice by focused-ion-beam milling. **d**, Energy dispersive X-ray elemental analysis for the device in c. hBN appears in green; $SiN_x$ in yellow. Scale bar, 500 nm. **e**, High angle annular dark field scanning transmission electron microscope (HAADF STEM) image of the interface between hBN and Pd. The imaged area is indicated in d. Scale bar, 50 nm. **f**, HAADF STEM image of hBN planes, which is taken from the area indicated in e. vdW gaps appear in dark. Scale bar, 5 nm.

To investigate the possibility of permeation of thermal protons and deuterons – nuclei of hydrogen isotopes – through vdW gaps, we have fabricated devices as described in Fig. 1 and in Supplementary Information. Thin crystals of hBN, $MoS_2$ or graphite (typically ~500 nm thick) were placed to seal circular holes of radius $r \approx 10$ μm that were etched in freestanding silicon nitride ($SiN_x$) membranes. The assembly separated two compartments and was vacuum-tight (Supplementary Fig. 1). Then the suspended crystal was plasma-etched into an 'inverted cup' shape to provide interlayer channels connecting the compartments (Figs. 1a,b). The channels had length $L$ and their number was controlled by the height $h$ that determined the number of crystallographic planes exposed on both sides (Fig. 1).

The devices were coated with a thin (~50 nm) Pd film, which ensured a good interface for proton transport (Supplementary Information). Indeed, protons easily dissolve and diffuse in Pd[16,17] (also see below). The final assembly was covered on both sides with proton conducting polymer, Nafion[18], and proton-injecting electrodes were attached for electrical measurements (Supplementary Fig. 1). The devices were placed in a chamber with 10%-$H_2$-in-Ar atmosphere at 100% humidity, which ensured high proton and negligible electron conductivity of Nafion[18]. If a voltage was applied between the electrodes, the protons injected into Nafion diffused through it and the Pd film before encountering the vdW crystal under investigation (inset of Fig. 2a). Therefore, a finite electrical current through the closed circuit was an unequivocal indication that hydrogen permeated through the vdW crystals. Our measurements could not distinguish whether it permeated in the form of charged protons/deuterons or neutral atoms because, in principle, a charged proton can acquire an electron from the environment or the 2D crystal as it transports through. Both interpretations are possible, as shown for hydrogen transport through Pd films[19,20]. Therefore, without loss of generality, we discuss our results below in terms of proton/deuteron transport. Note that molecular hydrogen ($H_2$) cannot permeate through the vdW crystals and this was also ruled out experimentally (see "Conductance measurements" in Supplementary Information).

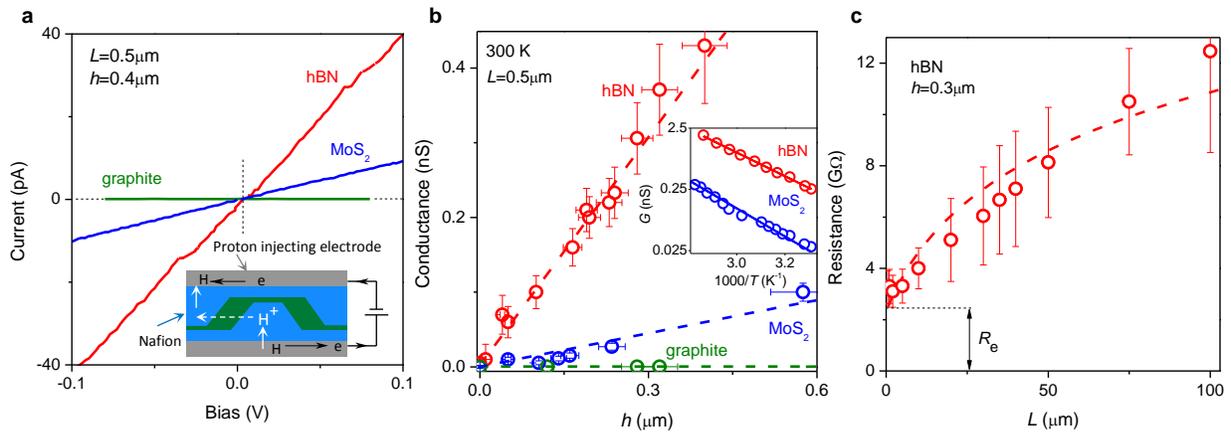

**Figure 2| Proton permeation for different devices and geometries. a**, Examples of *I-V* characteristics for devices made from graphite, hBN and $MoS_2$ with the same *L* and *h*. Inset, schematic of the devices and used circuitry. The etched vdW crystal (green) is covered with Nafion (blue) and the electrodes are shown in grey. **b**, Conductance *G=I/V* as a function of *h* for our shortest devices (*L* = 0.5 μm) for all three materials. Dashed lines, best linear fits. Inset, Arrhenius plots for hBN and $MoS_2$ devices with *L* = 0.5 μm, *h* = 0.25 μm. The horizontal error bars indicate uncertainty in *h* arising from the device fabrication procedure; the vertical ones are S.D. in the measured signal **c**, Dependence of the resistance 1/*G* on *L* for a fixed *h* for our hBN devices. Dashed curve, best fit to Ohm's law in the disk geometry. $R_e$ is the entry resistance in the limit of short *L*.

Fig. 2a shows typical current-voltage characteristics for our devices. These revealed a marked difference in the ability of protons to permeate through different layered crystals. In the case of hBN and $MoS_2$, the measured current *I* varied linearly with bias *V*, yielding the proton conductance *G=I/V*. For a fixed value of *L*, *G* was found to scale linearly with *h*, that is, the number of interlayer channels involved, as expected for our device geometry (Fig. 2b). The conductance values for hBN were an order of magnitude higher than for $MoS_2$. On the other hand, within our accuracy limited by leakage currents of ~0.1 pA, devices made from graphite exhibited no discernable conductivity for any $L$ and $h$. We emphasize that in all cases the finite $G$ could be attributed only to transport through the interlayer spacing. Indeed, reference devices made from hBN and $MoS_2$ but without etching ($h \equiv 0$), so that no interlayer channels were provided to connect the two compartments, exhibited no discernable conductivity. The latter observation also implies that vdW crystals are highly anisotropic with respect to proton transport and allow little permeation perpendicular to their atomic layers, in agreement with the previous reports[21,22]. As another reference, we fabricated samples where the two compartments were separated by a thin (~50 nm) Pd film but no crystal was placed in between. The measured $G$ was ~100 times higher than that of our most conductive hBN devices ($L$ = 0.5 μm and large $h$). This proves that Nafion and Pd layers contributed little to the measured overall resistance.

The measurements in Figs 2a,b clearly show that protons permeate easier along atomic planes of hBN than $MoS_2$. This is consistent with the activation energies found for the two materials. Indeed, the resistance of our devices exhibited an Arrhenius-type behavior: $R = 1/G \propto \exp(E/kT)$, where the activation energy $E$ was found to be 0.45±0.04 and 0.6±0.04 eV for hBN and $MoS_2$, respectively (inset of Fig. 2b and Supplementary Fig. 2).

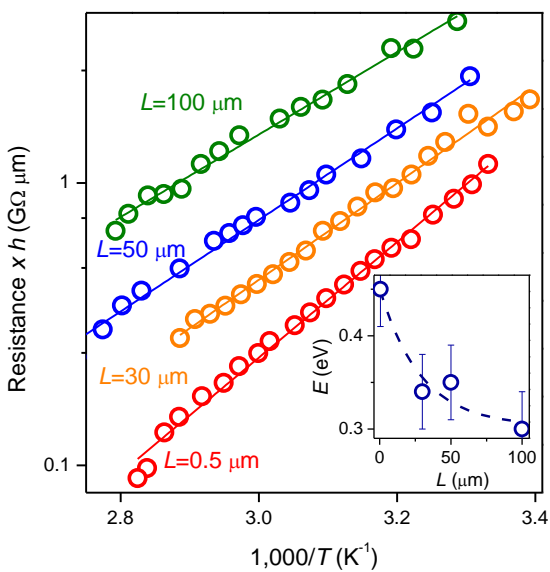

**Figure 3| Dependence of activation energy on channel length**. Arrhenius plots for hBN devices with different $L$. For consistency, the plotted resistance $R$ (in log scale) was normalized by $h$. Solid lines, best Arrhenius fits. Inset, activation energies $E$ for the data in the main panel. Error bars, S.D. Dashed curve, guide to the eye. Note that devices with different $h$ but same $L$ exhibited the same $T$ dependences, as expected (Supplementary Fig. 2).

Focusing on our most permeable material, hBN, we studied the dependence of its proton conductance on the channel length $L$. Figure 2c shows that the measured resistance $R$ increased only by a factor of ~4 when $L$ was increased 200 times. Crucially, unlike the $h$ dependence in Fig. 2b, the $L$ dependence did not extrapolate to zero in the limit of short $L$ (Fig. 2c). This behavior clearly indicates that protons experience an entry barrier that gives rise to the entry resistance, $R_e$, which dominates the permeation rate at short $L$. Given the 'entry' area involved in proton permeation, $2\pi rh$, the barrier can be described by an areal resistivity $\rho_e = \pi rh R_e$. For hBN at room $T$, extrapolating our resistance data in Fig. 2c to $L$=0 yields $\rho_e$ ~20 GΩ μm². For longer devices, the overall resistance has an additional contribution, $R_d$, associated with proton transport along the interlayer spacing. For our disk geometry, $R_d$ is expected to depend logarithmically on $L$ (dashed curve in Fig. 2c). Analysis of this dependence, described in Supplementary Information, yields the in-plane resistivity for protons inside hBN, $\rho_d \approx$ 7.5 GΩ μm. Further information about the entry and in-plane permeation was gained by measuring their activation energies: $E_e$ and $E_d$, respectively. Fig. 2c shows that for $L \approx 0.5$ μm, the device resistance was completely dominated by $R_e$. Therefore, the activation energy $E$ = 0.45±0.04 eV reported in the inset of Fig. 3 is *de facto* $E_e$. On the contrary, resistances of long devices ($L \approx 100$ μm) are dominated by $R_d$. By measuring $T$ dependences for the latter devices and taking into account a small but finite $T$-dependent contribution arising from $\rho_e$, we obtained $E_d$ = 0.26±0.04 eV (see Figure 3 and "Activation energies" in Supplementary Information).

What are the physical mechanisms governing the entry and in-plane resistivities? To answer this question we carried out isotope-effect experiments in which protons were substituted with deuterons following the recipe described in ref. 22 (for details, see "Conductance measurements" in Supplementary Information). Let us discuss first the in-plane transport. We repeated the $L$-dependence measurements of Fig. 2c with hBN but for deuterons. These showed that, for the same device, the in-plane resistivity $\rho_d^D$ was notably higher than $\rho_d^H$ (Fig. 4a). This is consistent with the diffusion of classical particles, in which case the diffusion rate is expected to be inversely proportional to $\sqrt{m}$. Within our data scatter, this factor fully accounts for the observed difference between $\rho_d^D$ and $\rho_d^H$. The classical interpretation is also consistent with our density functional theory calculations that yielded a barrier of ~0.2 eV, in good agreement with the observed $E_d \approx 0.26$ eV (Supplementary Information). Within this model, it is straightforward to understand the lower proton conductivity through MoS$_2$ and its complete absence for graphite. Indeed, previous experiments and calculations reported large activation energies for diffusion of atomic hydrogen in graphite[23,24] (~ 0.4-1.1 eV), which should translate into exponentially smaller permeation rates. The diffusion barrier is predicted to be considerably smaller for MoS$_2$[25], in agreement with our experiment.

It is instructive to compare the diffusion constant for protons in hBN, $D_{hBN}$, with those in other diffusion processes. From the found value $E_e$, we can estimate[4] the concentration of protons in hBN as $n_{hBN} \approx n_{PdH}$ exp($-E_e/kT$) where $n_{PdH}$ is the proton concentration in the Pd film. Because Pd is immersed in a proton-conducting medium (Nafion in our case), $n_{PdH}$ is expected to rapidly reach its known saturation value of ~1% atomic fraction[17]. This yields $n_{PdH} \approx 10^{21}$ cm$^{-3}$ and hence $n_{hBN} \approx 10^{13}$ – $10^{14}$ cm$^{-3}$. The main uncertainty in estimating $n_{hBN}$ arises from our limited accuracy (±40 meV) of determining $E_e$. The above estimate yields $D_{hBN} = kT/(\rho_d n_{hBN} e^2) \approx 10^{-4}$ - $10^{-3}$ cm² s$^{-1}$, where $e$ is the elementary charge. Such fast diffusion is consistent with the low activation energy for in-plane proton transport in hBN. In fact, $D_{hBN}$ is three orders of magnitude higher than typical diffusion constants for the intercalation of Li in graphite[26], a process utilized in Li batteries, and comparable or even faster than the diffusion constant of thermal protons in water[27] (see "Diffusion constant" in Supplementary Information).

Now we turn to the mechanism behind the entry barrier. Using our shortest devices ($L \approx 0.5$ μm), in which the entry resistance dominated, we found that $\rho_e^D$ in hBN was ~40% smaller than $\rho_e^H$ for protons (Fig. 4b): $\rho_e^D < \rho_e^H$. The same isotope effect was found for our MoS$_2$ devices, although in this case the difference was ~20%. This isotope effect shows that deuterons face a lower entry barrier than protons and is in stark contrast with the one observed for the in-plane transport where $\rho_d^D > \rho_d^H$. The substantial difference in entry barriers can be attributed to quantum sieving[5] through the vdW gaps at the exposed hBN edges.

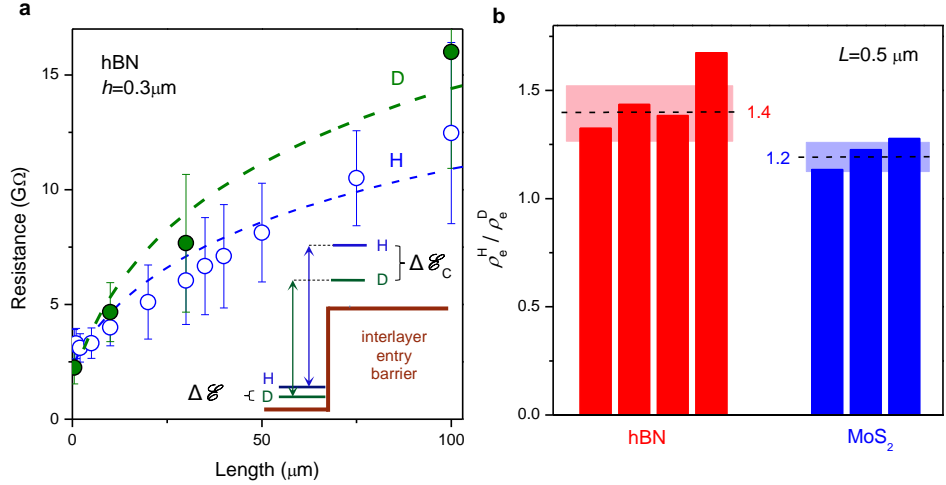

**Figure 4 | Isotope effect for interlayer transport. a**, Proton and deuteron resistance in hBN devices with $L$ ranging from 0.5 to 100 μm and $h$ = 0.3 μm. Dashed curves, best fits to Ohm's law in the 'inverted cup' geometry. Note that, despite the relatively large error bars, the data could also fit with a sublinear but quicker than logarithmic dependence. This would imply a somewhat smaller $\rho_d$, but would not influence other conclusions. Inset, schematic of the entry mechanism. The classical energy barrier for entry into the vdW gap is shown in brown. This barrier is modified by zero-point fluctuations for protons (H) and deuterons (D). The corresponding energy levels are indicated by blue and green horizontal lines, respectively. Inside the vdW crystal, strong confinement raises the effective barrier but differently for the two isotopes. The difference in their energies is $\Delta \mathscr{E}_C$. The barrier is also modified by zero-point energies inside Pd, which split the isotope levels by $\Delta \mathscr{E}_{Pd}$. **b**, Ratio of the entry resistances for protons and deuterons. hBN (red) and MoS$_2$ (blue) devices with our shortest $L$ = 0.5 μm. Each bar represents a different device. Dashed lines, average resistance ratio for each of the materials; shaded areas, standard errors.

The data in Fig. 4b allow us to estimate the effective width $W$ of these gaps. Due to the crystallographic confinement, the energy of both protons and deuterons should be shifted by $\mathscr{E}_C^H$ and $\mathscr{E}_C^D$, respectively (inset of Fig. 4a). Assuming a square-well confinement potential, we relate its width $W$ to the zero-point energy $\mathscr{E}_C^H$ as

$$W = h\,(8m^H \mathscr{E}_C^H)^{-1/2} \qquad (1)$$

where $m^H$ is the proton mass. An equivalent formula applies for deuterium. To extract $\mathscr{E}_C^H$ from our experimental data, we note first that the ratio $\rho_e^H/\rho_e^D$ is related to the entry barriers for protons ($E_e^H$) and deuterons ($E_e^D$) through the expression

$$\rho_e^H/\rho_e^D = \sqrt{(m^H/m^D)} \times \exp[\Delta E_e / k_B T] \qquad (2)$$

where $\Delta E_e = E_e^H - E_e^D$. Second, it was previously reported that protons and deuterons in Pd also exhibit a noticeable isotope shift[28,29] $\Delta \mathscr{E}_{Pd} = \mathscr{E}_{Pd}^H - \mathscr{E}_{Pd}^D \approx 20$ meV (inset of Fig. 4a). This shift effectively reduces the quantum sieving effect in our devices as

$$\Delta E_e = \Delta \mathscr{E}_C - \Delta \mathscr{E}_{Pd} \quad (3)$$

where $\Delta \mathscr{E}_C = \mathscr{E}_C^H - \mathscr{E}_C^D$. Using the above formulas, we obtain $\mathscr{E}_C^H = (1 - m^H/m^D)^{-1} \Delta \mathscr{E}_C = 2(\Delta E_e + \Delta \mathscr{E}_{Pd}) \approx 76$ meV for hBN and $\approx 68$ meV for MoS$_2$. From the found confinement energy, we estimate the width $W$ of the vdW gaps as $\approx 0.52$ and $0.55$ Å for hBN and MoS$_2$, respectively. These gaps are notably narrower than the de Broglie wavelenth of thermal deuterons ($\lambda_B^D \approx 1.02$ Å) and up to ~3 times narrower than that for protons ($\lambda_B^H \approx 1.45$ Å). This is consistent with both the large entry resistance $\rho_e$ for both isotopes and its higher value for protons than deuterons, as observed experimentally.

In conclusion, we demonstrate that interlayer gaps seen in TEM images of vdW materials (e.g., Fig. 1f) provide angstrom-scale channels that can distinguish between atoms and molecules with different de Broglie wavelengths. Furthermore, protons and deuterons are found to permeate through hBN and MoS$_2$ remarkably quickly. So, based on our findings, it is reasonable to expect that there are many other vdW crystals that exhibit substantial in-plane proton conductivity at room temperature. The unexpectedly fast diffusion of protons along the interlayer channels probably involves quantum effects, similar to the case of proton transport in water[30]. Finally, it would be interesting to investigate if vdW crystals can be combined with other materials considered to be promising for hydrogen sieving (for example, Pd films) to improve isotope separation technologies.

# Transport of hydrogen isotopes through interlayer spacing in van der Waals crystals
S. Hu, K. Gopinadhan, A. Rakowski, M. Neek-Amal, T. Heine, I. V. Grigorieva, S. J. Haigh, F. M. Peeters, A. K. Geim, M. Lozada-Hidalgo

**Device fabrication**

The device fabrication process is depicted in Supplementary Fig. 1. It starts by etching 20 μm through holes into SiN$_x$ membranes (see reference 1 for further details). Then, crystals of hBN, graphite or MoS$_2$ of initial thickness $h_0 \approx$ 500 nm were prepared by mechanical exfoliation and transferred over the SiN$_x$ through holes[1]. To form in-plane channels for hydrogen permeation, first, e-beam lithography and reactive ion etching were used from the top to form a ring-like structure of height $h_1$ (see Supplementary Fig. 1b). Second, the crystals were etched from the back to form a circular cavity of depth $h_2$. In this way, a channel with height $h=h_1+h_2-h_0$ and length $L$ is formed. The dimensions of each final structure were measured with AFM. This revealed that our fabrication procedures can control $h$ within an error of $\pm 10\%$ and $L$ within $\pm 5\%$. Once the above fabrication was completed, we deposited ≈20-50 nm of Pd with a Cr sublayer (for adhesion) via ebeam evaporation on both sides. Next, Nafion solution (5%, 1,100 equiv. weight) was drop cast and porous carbon electrodes with Pt catalyst were mechanically attached on both sides. The whole device was baked in a water-saturated environment to both crosslink the polymer and improve electrical contacts.

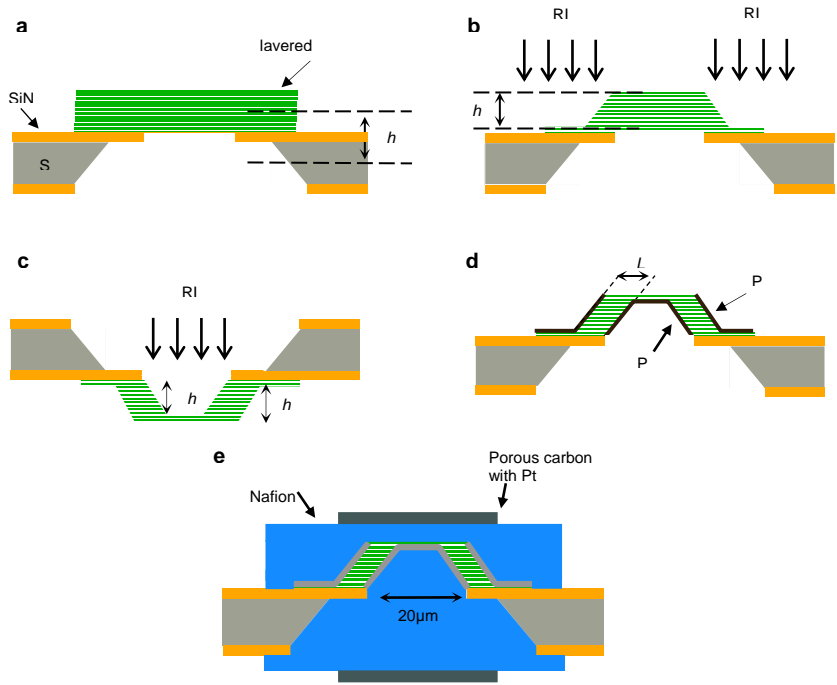

**Supplementary Figure 1| Schematic of our fabrication procedures. a,** Crystal transfer over the hole in SiN$_x$. **b,** Crystal etching from the top. **c,** Etching on the opposite side. **d,** Pd film deposition on both sides. **e,** Sample sandwiched between Nafion and PdH$_x$ electrodes.

9# Transport of hydrogen isotopes through interlayer spacing in van der Waals crystals

S. Hu, K. Gopinadhan, A. Rakowski, M. Neek-Amal, T. Heine, I. V. Grigorieva, S. J. Haigh, F. M. Peeters, A. K. Geim, M. Lozada-Hidalgo

**Device fabrication**

The device fabrication process is depicted in Supplementary Fig. 1. It starts by etching 20 μm through holes into SiN$_x$ membranes (see reference 1 for further details). Then, crystals of hBN, graphite or MoS$_2$ of initial thickness $h_0 \approx$ 500 nm were prepared by mechanical exfoliation and transferred over the SiN$_x$ through holes[1]. To form in-plane channels for hydrogen permeation, first, e-beam lithography and reactive ion etching were used from the top to form a ring-like structure of height $h_1$ (see Supplementary Fig. 1b). Second, the crystals were etched from the back to form a circular cavity of depth $h_2$. In this way, a channel with height $h=h_1+h_2-h_0$ and length $L$ is formed. The dimensions of each final structure were measured with AFM. This revealed that our fabrication procedures can control $h$ within an error of $\pm 10\%$ and $L$ within $\pm 5\%$. Once the above fabrication was completed, we deposited ≈20-50 nm of Pd with a Cr sublayer (for adhesion) via ebeam evaporation on both sides. Next, Nafion solution (5%, 1,100 equiv. weight) was drop cast and porous carbon electrodes with Pt catalyst were mechanically attached on both sides. The whole device was baked in a water-saturated environment to both crosslink the polymer and improve electrical contacts.

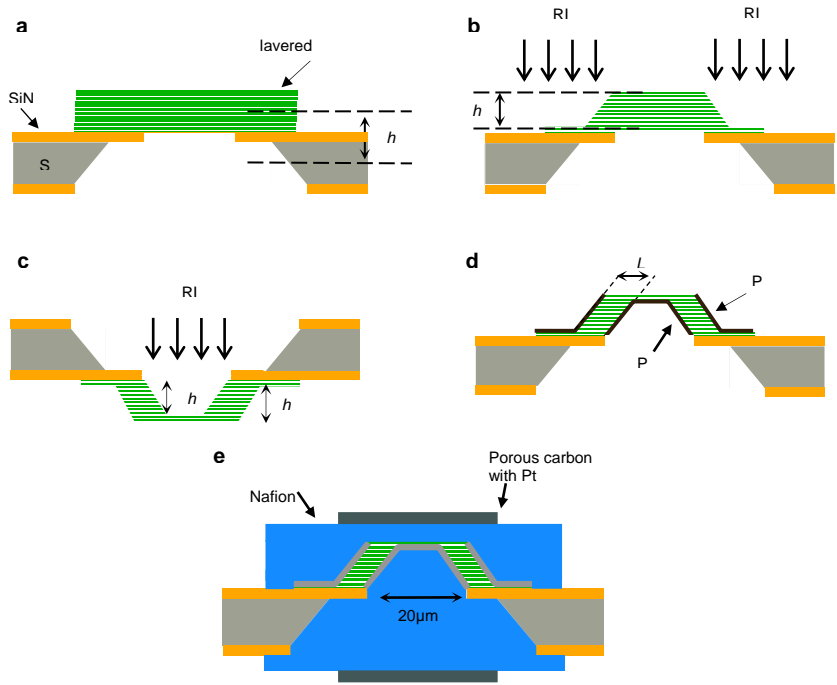

**Supplementary Figure 1| Schematic of our fabrication procedures. a,** Crystal transfer over the hole in SiN$_x$. **b,** Crystal etching from the top. **c,** Etching on the opposite side. **d,** Pd film deposition on both sides. **e,** Sample sandwiched between Nafion and PdH$_x$ electrodes.

**SEM, FIB and STEM characterization**

To obtain the cross sectional SEM image shown in Fig. 1 of the main text, a dual beam FIB-SEM system (Helios 660, FEI) with a focused ion beam (FIB) and a field emission gun was used. Prior to ion milling, a protective layer of amorphous carbon/Au-Pd (7nm/2nm) was sputtered onto the wafer, (Q150T, Quorum Technologies). The region of interest was identified using SEM, and a Pt protective layer (~2μm) was deposited using the FIB column. A trench was then milled (Ga$^+$ beam 30 kV, 2.4 nA ion current) adjacent to the Pt layer to reveal a cross section of the membrane structure. After further ion beam polishing back scattered electron images were acquired of the fresh surface (5 kV, 0.2 nA electron current). By removing further material from the exposed cross section, it was possible to confirm the intact 3D structure of the device.

Cross-sectional samples suitable for imaging with a scanning TEM were obtained in the FIB using the lift-out preparation technique[2]. Prior to ion milling, both sides of the device were protected by a layer of amorphous carbon/Au-Pd coating (7nm/2nm on top and 1μm/0.7μm on the back side for sample mechanical stability). A further 2 μm of Pt was deposited in situ on the topside region of interest, using the FIB column. A lamella cross sectional slice (1.5 – 2 μm thick) through the membrane was obtained by Ga$^+$ ion milling and transferred to a Cu three post TEM support grid using a nanomanipulator (EasyLift$^{TM}$, FEI). The lamella was then thinned until electron-transparent (< 50 nm) in the region of interest, using successively low beam energies and currents (30 – 2 kV, 240 – 10 pA)[3]. High-resolution STEM images and energy dispersive X-ray spectroscopy (EDXS) spectrum images were obtained using a probe side aberration corrected Titan G2 80 – 200 with a X-FEG electron source fitted with detector Super-X EDXS. High angle annular dark field (HAADF)-STEM images were acquired at 200 kV, with a convergence angle of 21 mrad, an inner angle of 54 mrad and a probe current of 180 pA. The lamella was aligned using the fringes in the boron nitride. EDXS data were quantified using a Cliff Lorimer analysis in the Esprit software version 1.9 (Bruker, USA).

**Conductance measurements**

To characterize the sensitivity of our experimental setup, we first measured an hBN device etched from both top and bottom directions, but the etching was planned so that no channels were formed ($h_1+h_2<h_0$). At room temperature, a parasitic conductance of <1pS was found in this case due to leakage currents along the Si wafer surface under the high humidity environment. This value represents the lower bound of our measurement sensitivity. Then, to characterize the upper bound, we fabricated devices in the same way but without a crystal (still with ~20-50 nm Pd film). These showed a room-temperature conductance of ~10 nS. Within these two values, then, we could measure the transport characteristics of our devices, ignoring the leakage and contact resistances.

Note that the Pd film is important for our measurements. Without it, the conductance was found to be badly reproducible even for devices having the same *h* and *L*. We attribute this to poor contact between our nanometer-scale structures and the drop-cast polymer (Nafion). In contrast, the evaporated Pd smoothly conforms to structured surfaces of the layered crystals[4], providing good contact. We have confirmed that the Pd film does not limit proton transport through our devices. Indeed, devices with such films but without vdW crystals exhibited a conductance ~100 times higher than any studied hBN device, as described above. Furthermore, we varied the Pd films thickness from ~8 up to ~50 nm in similar hBN devices and found the same conductance regardless of the film thickness.

It is also important to notice that molecular hydrogen cannot diffuse through the vdW gaps. We can rule this out both theoretically and experimentally. For hydrogen molecules to diffuse through, it would be necessary to expand the crystal lattice[5]. This would require an energy equivalent to the cleavage energy that is ~60 meV per atom[6,7]. But this value is comparable to the energy necessary to combine two hydrogen atoms in Pd or Pt to form a hydrogen molecule[8] (~100 meV). Therefore, the van der Waals forces that hold the layers together would prevent the formation of hydrogen molecules. From an experimental perspective, transport of molecular hydrogen would result in bubbles and quickly destroy the interface between the thin Pd layer and our vdW crystals, as reported previously[1,9]. In this work, we did not observe such bubbles, and our devices can be measured for several days without degradation.

For isotope-effect measurements, we substituted protons (H) with deuterons (D) as previously reported[9,10]. Typically, the same device was measured with both protons and deuterons, which allowed higher accuracy of determining the ratio $\rho_e^H/\rho_e^D$. To this end, the chamber was filled with $D_2$ gas instead of $H_2$ and $H_2O$ vapor was substituted with $D_2O$ vapor. This method is known[11,12] to be able to substitute protons for deuterons in Nafion. Indeed, we verified that only ≈1% of protons remain in our devices under these conditions[9]. Notwithstanding, to ensure the reproducibility of the measurement, the device was cycled several times between H and D. For a given device, we always observed the same $\rho_e^H/\rho_e^D$ ratio within an accuracy of approximately 10%.

At this point, it is instructive to estimate the possible effect of the 1% remnant protons in Nafion during deuteron transport measurements. We note that, for a given H-D ratio in Nafion, the ratio of H-D ratio inside palladium will be given by its room temperature proton-deuteron separation factor[13,14]: $\alpha_{HD}$ = $[H/D]_{Pd}/[H/D]_{Nafion}$ ≈ 2 where $[H/D]_{Pd(Nafion)}$ is the ratio of atomic fractions of protons and deuterons in palladium or Nafion. Therefore, we estimate that after the H-D substitution only ≈2% of the atoms dissolved in palladium are remnant protons. Hence, the measured entry resistivity for deuterons in our devices, $\rho_e^D$, will be given by $\rho_e^D$ = 0.98 $\rho_e^{D*}$ + 0.02 $\rho_e^{H*}$, where $\rho_e^{D*}$ refers to the actual – rather than measured – deuterium entry resistivity and $\rho_e^{H*}$ is the entry resistivity associated with residual protons. Because the electrical measurements reported in the main text yielded $\rho_e^H/\rho_e^D$ ≈ 1.4, the above consideration yields $\rho_e^{H*}/\rho_e^{D*}$ ≈ 1.41. Therefore, the correction due to H contamination of D-Nafion is relatively small and within the statistical error of our measurements, as shown in Fig. 4 of the main text.

**Resistance analysis**

According to Ohm's law, the resistance across the space between two cylinders of height $h$ and radii $r$ and $r+L$ is given by

$$R_d = \int_r^{r+L} \rho_d/(2\pi rh)dr = [\rho_d/(2\pi h)] \ln(1+L/r) \qquad (1)$$

where $\rho_d$ is the bulk resistivity of the material. This formula yields of course that, in the limit of small $L$, $R_d$ approaches zero. Accordingly, to model the $L$-dependence observed for our devices, an additional resistance term – associated with entry into the interlayer space – must be included. The latter term is given by

$$R_e = [\rho_e/(2\pi h)] \{(1/r) + [1/(r+L)]\} \qquad (2)$$

where $\rho_e$ is the areal entry resistivity. The total resistivity is given by $R = R_e + R_d$. In the limit of small $L$, $R \approx R_e = \rho_e/\pi rh \propto \rho_e(T)$ where $T$ is the temperature. On the other hand, in the limit of large $L$, $R \approx R_d = [\rho_d/(2\pi h)] \ln(1+L/r) \propto \rho_d(T)$.

**Activation energy**

The plotted dependences of proton conductance on the device geometry ($h$ and $L$) in Figs 2b,c show that our measurements are highly reproducible with a relatively small data scatter. The horizontal error bars are mostly due to our etching accuracy whereas the vertical ones due to electrical noise and sweeping hysteresis (large time constants, $RC$). Taking into account the exponential dependence of proton conductivity on temperature, the reproducibility also implies that the activation energy $E$ should not vary much between similar devices. Nonetheless, we investigated the reproducibility of $E$ directly. Supplementary Fig. 2 shows Arrhenius plots of our conductance measurements for both hBN and $MoS_2$. The data show that the extracted $E$ is reproducible within ~40 meV. Furthermore, it shows that devices with the same $L$ but different $h$ exhibit the same activation energy, as expected.

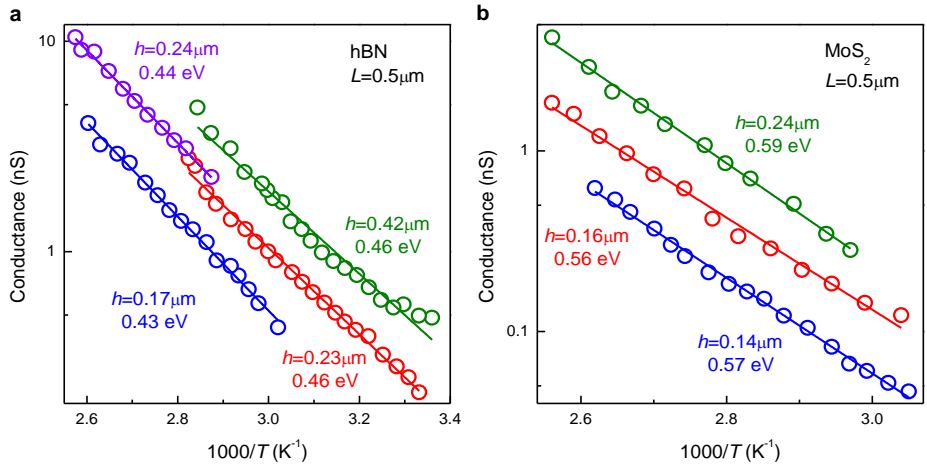

**Supplementary Figure 2 | Reproducibility of activation energies for different devices. a,** Temperature dependence for four hBN devices with different $h$ but same $L$ = 0.5 μm. **b,** Temperature dependence of conductance for four $MoS_2$ devices. No changes in $E$ were observed for different $h$ within our accuracy.

We investigated the activation energies in the limit of short and long $L$ to extract the in-plane and entry activation energies: $E_e$ and $E_d$, respectively. The entry process is straight forward to analyze because the resistance of devices with $L \approx 0.5$ μm is practically indistinguishable from the entry resistance $R_e$. On the other hand, devices with $L \approx 100$ μm can be used to extract the in-plane activation energy as follows. Since both total and entry resistances ($R$ and $R_e$, respectively) are activated, the in-plane resistance as a function of $T$ is $R_d(T)=R(T)-R_e(T)$. Furthermore, $R_e(T)$ is given by the resistance of devices with $L \approx 0.5$ μm. Using this subtraction for $L \approx 100$ μm and fitting the result with the Arrhenius equation yielded $E_d=0.26\pm0.04$ eV. *A posteriori*, this analysis shows that the resistance of devices with $L \approx 100$ μm is a close approximation to $R_d$. The analysis yielded only a small (~10%) correction with respect to the procedure that ignores the entry resistance.

**Diffusion constant**

It is instructive to estimate the diffusion constant of protons in our vdW crystals. To this end, resistivity is related to the diffusion constant as $D_{hBN}= (1/\rho_d) k_BT/(n_{hBN}e^2)$ where $n_{hBN}$ is the proton concentration in hBN, $e$ is the elementary charge, $k$ the Boltzmann constant and $T$ the temperature. We estimate the proton concentration in hBN as $n_{hBN} \approx \exp(-E_e/kT)xn_{Pd}$. Here $n_{Pd}$ is the concentration of atoms in PdH$_x$, $x$ is the fraction of protons in PdH$_x$, and $E_e$ is the activation energy for proton entry into hBN. Since[15] $x$ is typically ~1%, we neglect changes in density and molar weight in pure palladium due to proton absorption and estimate the density of atoms in PdH$_x$ from the values known for pure Pd as $n_{Pd}$ = density/molar weight $\approx$ 12 g cm$^{-3}$/106 g mol$^{-1}$ ~$10^{23}$ cm$^{-3}$. Hence, from the measured entry activation energy ($E_e \approx$ 0.4 eV), we estimate $n_{hBN}$ to be of the order of $10^{13} – 10^{14}$ cm$^{-3}$. The main uncertainty here arises from the statistic error of $\pm$0.04 eV in our measurements of $E_e$. From these values of $n_{hBN}$ and using the measured $\rho_d$ we estimate $D_{hBN}$ as ~$10^{-4} – 10^{-3}$ cm$^2$ s$^{-1}$.

**Density functional theory calculations**

We studied the transport of a hydrogen atom along the interlayer space of hBN using density functional theory (DFT). We performed *ab initio* calculations for an AA' stacked hBN bilayer (interlayer distance $d$ = 3.34 Å) consisting of 190 atoms passivated at the edges with 50 hydrogen atoms. Adding more hBN layers did not introduce significant difference to our results. The calculations were performed with the software package GAUSSIAN (G09 package) employing B3LYP/6-31G* functional-basis[16,17] and the van der Waals interactions were introduced using Grimme's empirical corrections[18].

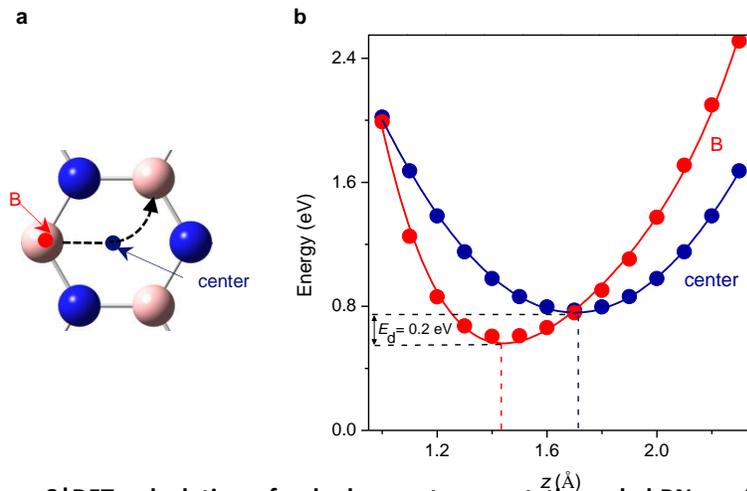

**Supplementary Figure 3|DFT calculations for hydrogen transport through hBN**. **a**, Diffusion trajectory for a hydrogen atom is indicated with the black arrow. Blue spheres, nitrogen atoms; pink, boron. **b**, Energy of atomic hydrogen confined between hBN layers for two fixed positions in the *x-y* plane.

To study the transport along hBN layers we calculated the adsorption energy of a hydrogen atom at fixed points in the *x-y* plain in hBN but for all possible positions along the *z* axis. These calculations, therefore, reveal the minimum energy that the hydrogen atom can have at a fixed point in the *x-y* plain if the *z* coordinate is allowed to relax. From this analysis, the diffusion path of atomic hydrogen and the energy barriers it faces along such a path can be inferred. The calculations were made as follows.

We calculated the energy of a hydrogen atom fixed above ($z > 0$) an arbitrary boron atom (B) in the bottom layer, in the center of the hexagonal ring and above a nitrogen (N) atom (Supplementary Fig. 3a). All these calculations are done as a function of *z*. Our results are illustrated in Supplementary Fig. 3b. The minimum energy is found above boron, $E_B$ = 0.56 eV and a distance of ~1.4 Å. The second lowest minimum occurs at the hexagon center with energy $E_C$ =0.77 eV when the H-atom is exactly in the middle between the two BN sheets (Supplementary Fig. 3b). The N-site, on the other hand, is endothermic with a repulsive energy of ~0.57eV. These results suggest the diffusion path for hydrogen: from a B atom, through the center of the hexagonal ring – as far as possible from the N atom – and then on to another B atom. Along that path the hydrogen atom moves along the z axis by 0.3 Å and faces an energy barrier $E_d = E_C - E_B \approx 0.2$ eV, in good agreement with the experiment.

Note that experimentally we cannot differentiate between transport of charged protons and neutral hydrogen atoms. Given that the measurable quantity is the proton current through our devices, we have interpreted our results in the main text as proton permeation. However, the DFT analysis is much simpler if atomic hydrogen is considered. Nonetheless, we believe that analysis for protons analysis should yield close results for $E_d$.

**Estimating van der Waals gaps**

We estimated the width of the vdW gaps with respect to protons following the procedure described in the main text. The analysis relies on data for the isotope-effect in Pd from calorimetry and neutron scattering experiments[19-24]. Those measured zero point energies for protons and deuterons ($\mathscr{E}_{Pd}$) and reported $\Delta \mathscr{E}_{Pd} \approx$ 10-30 meV. The large variations are due to several factors. First, $\mathscr{E}_{Pd}$ values depend on the particular lattice site the proton is adsorbed in Pd[20] and, therefore, crystallinity is important. Second, the presence of other metals can reduce the zero point energy, as demonstrated for silver/palladium alloys[20,23]. These two factors could also be relevant for our experimental system because the e-beam evaporated Pd exhibits little crystallinity and we used a Cr sublayer for adhesion. Therefore, our estimates used the midpoint value $\Delta \mathscr{E}_{Pd} \approx$ 20 meV but the other possible values would result in relatively minor changes in the width *W* of the vdW gaps.

Furthermore, our model for estimating *W* has assumed a constant value of the entry barrier $E_e$ whereas the barrier obviously depends on the de Broglie wavelength $\lambda_B$ and, therefore, on *T*. It is expected that $E_e$ should in principle increase with decreasing temperature because $\lambda_B$ becomes increasingly larger with respect to the entry size *W*. The above assumption is justified by the fact that we extracted *W* from the isotope effect measured at a constant (room) temperature and changes in $E_e$ due to temperature do not play a role in these measurements. On the other hand, our analysis of $E_e$ did involve *T* dependent measurements. However, we varied *T* only by ±30 K from a midpoint value of 320 K. Accordingly, the de Broglie wavelength changed by ±5% in those experiments, and changes in the barrier height can be estimated to be ±20 meV. These values are well within our statistical error in the Arrhenius measurements. Hence, constant $E_e$ is a good approximation for analysis of our experiments.

**Supplementary list of references**